\documentclass[11pt]{article}

\usepackage{amsmath,amssymb}
\usepackage{slashed}
\usepackage{xcolor} 
\usepackage{graphicx}
\usepackage{mathtools}

\usepackage{amstext}
\usepackage{amsmath}
\usepackage{multirow}
\usepackage{hhline}
\usepackage{amssymb}
\usepackage{hyperref}
\usepackage[all]{hypcap}
\usepackage{xspace}
\usepackage{endnotes}
\usepackage{rotating} 
\usepackage{subfigure}
\usepackage{epstopdf}
\usepackage{graphicx}
\usepackage{color}
\usepackage[export]{adjustbox}
\usepackage{progressbar}
\usepackage{caption}
\usepackage{braket}

\usepackage{url}
\usepackage{cite}

\newcommand{\beq}{\begin{equation}}
\newcommand{\eeq}{\end{equation}}
\newcommand{\bea}{\begin{eqnarray}}
\newcommand{\eea}{\end{eqnarray}}

\newcommand{\fref}[1]{Fig.~\ref{fig:#1}} 
\newcommand{\eref}[1]{Eq.~\eqref{eq:#1}} 

\newcommand{\sref}[1]{Sec.~\ref{sec:#1}}

\newcommand{\tref}[1]{Table~\ref{tab:#1}}

\begin{document}

\begin{center}
\vspace*{5mm}

\vspace{.01cm}
{\LARGE
{\bf Breakdown of Effective Field Theory for a \\ \vspace{.3 cm} Gluon Initiated Resonance} 
} \\
\vspace{.7cm}
{\bf Alejandro de la Puente and Daniel Stolarski}

\vspace*{.3cm} 
Ottawa-Carleton Institute for Physics, Carleton University,\\ 1125 Colonel By Drive, Ottawa, Ontario K1S 5B6, Canada
\vspace*{.2cm} 

\end{center}

\vspace*{.5mm}
\begin{abstract}\noindent\normalsize
Gauge invariance dictates that a resonance produced from initial state gluons must be produced through a non-renormalizable operator or a loop process. Should such a resonance be discovered, uncovering the dynamics that give rise to its couplings to gluons will be crucial to understanding the nature of the new state. Here we study how the production of this resonance at high transverse momentum in association with one (or more) jets can be used to directly measure the scale of the operator or the mass of the particles in the loop. We use a 750 GeV diphoton resonance as an example application, and we study how the non-renormalizable operator case can be described by a slowly converging effective field theory (EFT) expansion with operators of dimension five and seven. We show that with O(100) events, one can put strong constraints on the scale of the EFT, particularly in theories with strong coupling.  We also compare the EFT analysis to that of a UV completion with vector-like quarks, and outline how the mass of said quarks could be measured.
\end{abstract}

\section{Introduction} \label{sec:intro} 

The excess in the diphoton spectrum observed by both ATLAS~\cite{ATLAS} and CMS~\cite{CMS:2015dxe} has received tremendous interest from the theory community (for a review, see~\cite{Strumia:2016wys}). Should this excess be a real state, the most common explanation in the literature is a spin-0 state coupled to the gluon and photon field strength tensors. If this is the case, then there must exist additional states beyond the Standard Model (SM)~\cite{Knapen:2015dap} because the couplings to the field strength tensors are non-renormalizable operators. If the additional states are sufficiently heavy, the dynamics of the new resonance, which in this work we call $\phi$, can be described by an effective field theory (EFT), as has been emphasized in~\cite{Berthier:2015vbb,Kamenik:2016tuv,Franceschini:2016gxv}. 

At sufficiently high energy, the effective field theory will no longer be a good description. If this new state emerges from a strongly coupled theory~\cite{Harigaya:2015ezk,Nakai:2015ptz,Franceschini:2015kwy,Molinaro:2015cwg,Bai:2015nbs,Belyaev:2015hgo,Bian:2015kjt,Craig:2015lra,Franzosi:2016wtl,Harigaya:2016pnu,Hong:2016uou,Redi:2016kip,Harigaya:2016eol,Kamenik:2016izk,Ko:2016sht,Foot:2016llc,Iwamoto:2016ral,Hamaguchi:2016umx,Matsuzaki:2016joz,Bai:2016vca}, as implied by the hints of it having a large width in the ATLAS data~\cite{ATLAS}, then one might expect EFT to break down at quite a low scale. In this work we investigate this possibility by looking at production of $\phi$ in association with one (or more) high $p_T$ jets so that the $\phi$ also has large transverse momentum. This kinematic configuration can resolve the dynamics that generate the higher dimensional operators in the EFT. 

This technique has been explored in the context of Higgs physics~\cite{Harlander:2013oja,Grojean:2013nya,Banfi:2013yoa,Azatov:2013xha,Bramante:2014gda,Schlaffer:2014osa,Buschmann:2014twa,Dawson:2014ora,Ghosh:2014wxa,Edezhath:2015lga,Dawson:2015gka,Langenegger:2015lra,Bishara:2016jga,Soreq:2016rae} to explore if there are contributions to the Higgs coupling to gluons in addition to those from the SM. The case of a new resonance is qualitatively different from the Higgs, however, because for the Higgs we know that the dominant contributor to the loop is the top quark, and most previous work focuses on searching for small additional effects. For a new resonance, especially if it has significant decays to $\gamma\gamma$, the top and other SM states must be subdominant contributions to the gluon loop, so we are here searching for the leading effects in the generation of that loop.  

An alternative way to look for breakdown of EFT is of course to directly observe new states. The technique we will describe is complementary to direct searches because it can unravel the nature of the interactions of any new states with the $\phi$. Furthermore, exploring associated production with jets is independent of decay modes of additional new states, so if those new states are somehow buried under background, they can still be uncovered with indirect techniques.

In this work we will show that the $p_T$ distribution of the $\phi$ can probe the breakdown of the EFT, or alternatively probe the details of the underlying UV completion. For concreteness, we will use the mass and cross section of the resonance hinted at here~\cite{ATLAS,CMS:2015dxe}, but we stress that this technique applies to any resonance coupled to gluon pairs. This follows from gauge invariance: there is no renormalizable operator one can write down that couples two gluons to one state, so new states must mediate this interaction. If the new states are heavy, one can describe the interaction via non-renormalizable operators, but if the new states are light then their full effects must be included. Should any such resonance be discovered, whether it be the 750 GeV diphoton resonance hinted by the 2015 data~\cite{ATLAS,CMS:2015dxe}, or some eventual discovery in the forthcoming runs of LHC or a future hadron collider, the work presented here will be a useful post-discovery tool in many BSM scenarios.

This work is organized as follows. In \sref{eft} we describe our effective field theory analysis of $\phi$ + jet production and show how considering dimension seven operators can significantly affect the $p_T$ distribution, especially at high $p_T$. Then we describe our statistical procedure to place a lower bound on the scale of the EFT, and show how strong of a lower bound one can place as a function of the total number of $\phi$ events. In \sref{vlq} we compare the EFT analysis to a UV completion with vector-like quarks (VLQ). We show that the two computations agree when the $p_T$ is low compared to the VLQ mass and disagree at high $p_T$ as expected. We then explore different features that appear in the spectrum with a genuine UV completion. Conclusions and descriptions of ways to improve and refine the analysis are given in \sref{conc}.

\section{Effective Field Theory Description} \label{sec:eft}

If the new states in addition to $\phi$ are sufficiently heavy, then production of a scalar (pseudo-scalar) $\phi$ at a hadron collider is dominated by the following dimension five operator:
\begin{equation}
\mathcal{O}_1 = \frac{g^{2}_{s}c_1}{\Lambda} \, \phi \, G^a_{\mu\nu} G^{a\,\mu\nu} \;\;\;\;
\left(\widetilde{\mathcal{O}}_1 = \frac{g^{2}_{s}\tilde{c}_1}{\Lambda} \, \phi \, G^a_{\mu\nu} \widetilde{G}^{a\,\mu\nu} \right) \;,
\label{eq:op5}
\end{equation}
where $g_s$ is the $SU(3)$ gauge coupling, $G^a_{\mu\nu} = \partial_\mu A_\nu^a - \partial_\nu A_\mu^a + g_s f^{abc} A_\mu^b A_\nu^c$ is the usual gluon field strength tensor, $c_1$ ($\tilde{c}_1$) is a dimensionless coupling constant and $\Lambda$ is the scale of the EFT. For the pseudo-scalar, 
$\widetilde{G}_{\mu\nu}^a=\frac{1}{2}\epsilon_{\mu\nu\rho\sigma}G^{a\rho\sigma}$ is the usual dual field strength tensor. 

At next-to-leading order in the EFT expansion, one expects dimension seven operators such as
\begin{equation}
\mathcal{O}_2 = \frac{g^{3}_{s}c_2 }{\Lambda^3} \, \phi \, G^{a\,\nu}_{\mu}\, G^{b\,\rho}_\nu \,G^{c\,\mu}_\rho \,f^{abc}\;\;\;\;
\left(\widetilde{\mathcal{O}}_2 = \frac{g^{3}_{s}\tilde{c}_2}{\Lambda^3} \, \phi \, \widetilde{G}^{a\,\nu}_{\mu}\, G^{b\,\rho}_\nu \,G^{c\,\mu}_\rho \,f^{abc} \right)\, 
\label{eq:op7}
\end{equation}
to become relevant, and we generically expect that $c_2$ is parametrically the same size as $c_1$. For processes at scales much lower than $\Lambda$, the operators in \eref{op7} will be subdominant relative to those in \eref{op5}. On the other hand, by probing processes at higher energies, one can see where the effects of the dimension seven operators become significant, and therefore measure the scale $\Lambda$. By looking at the structure of the operators in \eref{op7}, its clear that the process of $\phi$ + jet production is a way to access the high energy behaviour of dimension five and dimension seven operators, and that is what we study in this section.

We note that there are other dimension seven operators consistent with the SM gauge symmetry that lead to $\phi$ + jet production~\cite{Gracey:2002he,Neill:2009tn}, and these are studied in more detail for the Higgs in~\cite{Harlander:2013oja}. For a specific UV completion, one can compute the coefficients of all higher dimensional operators in terms of a few UV parameters. In this work we will, for simplicity, focus on the dimension 7 operators in \eref{op7} as a way to parameterize the breakdown of the EFT, and we leave a full study of all operators and interference effects to future work.

We simulate events using FeynRules~\cite{Alloul:2013bka} to create the model with a $\phi$ of mass of 750 GeV that couples to both the photon and gluon field strength tensors, implementing the operators in \eref{op5} and \eref{op7} for the gluons. We then use the Madgraph 5~\cite{Alwall:2011uj}, Pythia 6~\cite{Sjostrand:2006za}, Delphes~\cite{deFavereau:2013fsa} pipeline to simulate events at leading order. We also simulate the background using the same pipeline, and we apply a QCD K-factor of 2~\cite{Gehrmann:2013aga} to the background. We require that all events have at least one jet with $p_{T,j_{1}}>100$ GeV and pseudo rapidity, $|\eta_{j}|<5$. In addition, we require exactly two photons with $p_{T,\gamma}>20$ GeV and $|\eta_{\gamma}|<2.5$ and with an invariant mass in the range $680~\text{GeV} < m_{\gamma\gamma} < 820~\text{GeV}$.

As a proof of the principle, we show the $p_T$ spectrum of $\phi$ production in \fref{eft-spectra} for the case where dimension 7 operators are turned off (left figure) and turned on (right figure). Here we see that at high $p_T$ the effects of dimension seven operators become pronounced. Unfortunately, the spectrum is a rapidly falling function of $p_T$, so significant luminosity is required to see any effects for large $\Lambda$. We do know that substantially more data will be taken at the LHC, and we here quantify how much is needed to measure or exclude a given value of $\Lambda$.

\begin{figure}[tb]
\centering
\includegraphics[width=0.5\textwidth]{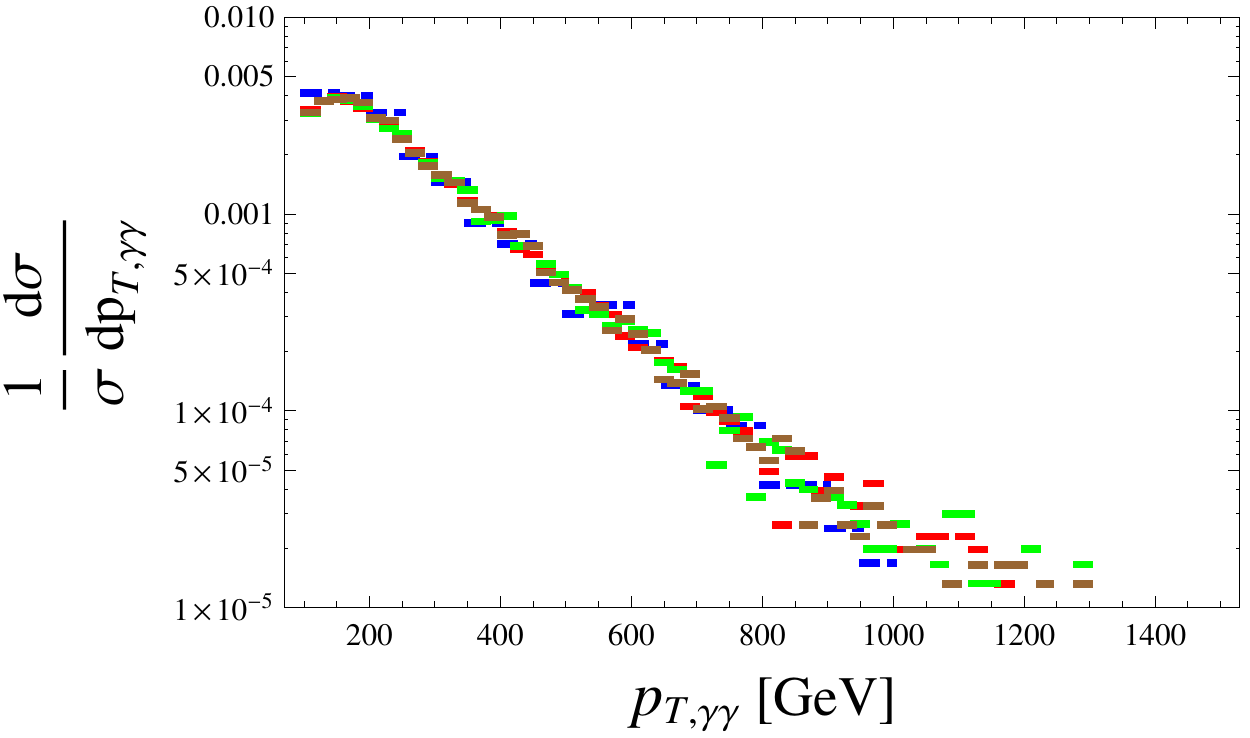}~~
\includegraphics[width=0.5\textwidth]{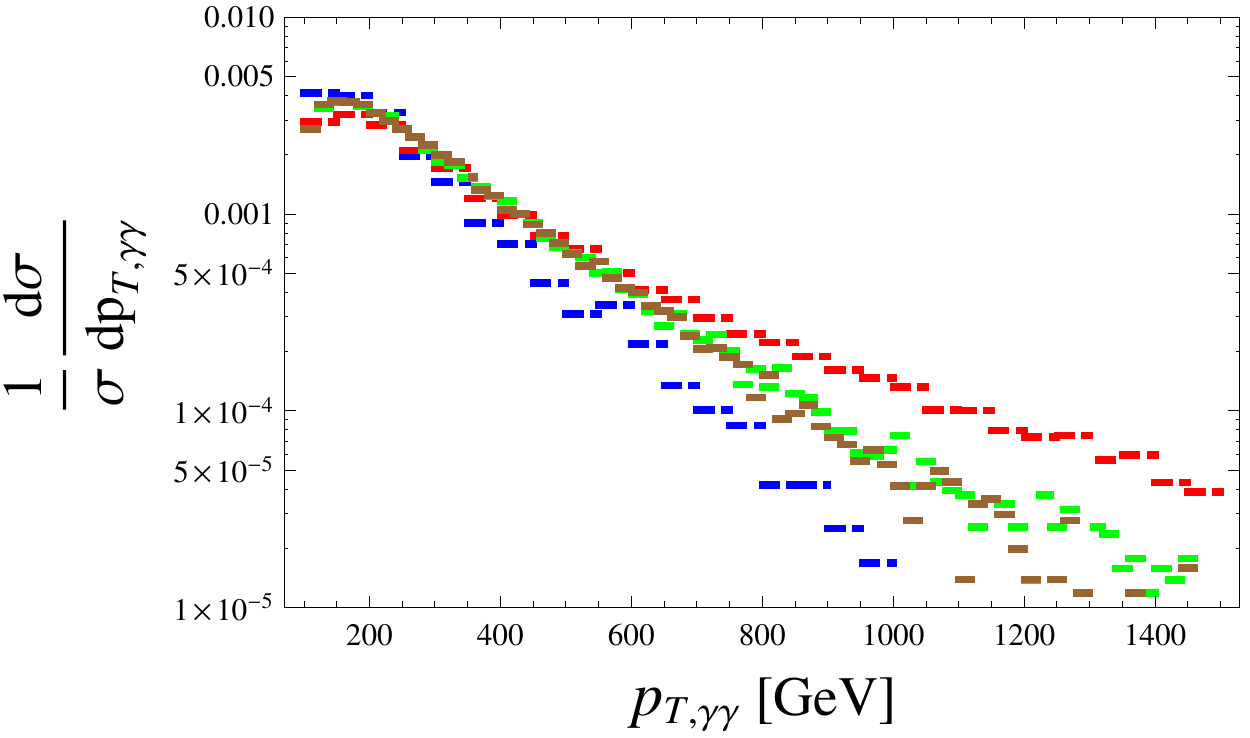}
\caption{ Normalized $p_T$ distribution of the scalar $\phi$ (decayed to $\gamma\gamma$) for the case where dimension 7 operators are turned off (left figure) and turned on (right figure). Here we take $c_1=c_2=1$, and $\Lambda = 750,~1250,~2000$ GeV in red, green, and brown, respectively. The background is depicted by the blue histogram. }
\label{fig:eft-spectra}
\end{figure}

The effects of dimension seven operators are expected to become important when the energy flowing into the vertex is of order $\Lambda$. This occurs when each of the $\phi$ and the jet carry energy of $\Lambda/2$. Here we use $p_T$ as a stand in for energy: we probe the existence of a given value of $\Lambda$ by placing a $p_T$ cut of $\Lambda/2$ on both the leading jet and the diphoton system, and then ask the total cross section when this cut is applied.  We compare the number of events that pass this cut as well as all the pre-selection cuts described above for a fixed $\Lambda$ to the number of events with $\Lambda = 10$ TeV to capture the effects of only including the dimension five operator. We also include continuum diphoton background in both samples, but we require $680 < m_{\gamma\gamma} < 820$ GeV. The cross sections for various representative values of $\Lambda$ with a $p_T$ cut of $\Lambda/2$ are shown in \tref{xsec}. 

\begin{table}[tb]
\centering
\begin{tabular}{|c|c|c|c|c|}
\hline
 & $\Lambda$ (GeV) & $\sigma$ (fb) & $\sigma_{bkg}$ (fb) & $\sigma_{\Lambda\to\infty}$ (fb)  \\ \hline\hline 
$\phi$ production (inclusive)  & - & 8 &  &\\ \hline\hline
$\phi$ production & 750 & $5.1\times10^{-1}$ &  $8.0\times10^{-1}$ & $3.6\times10^{-1}$ \\ 
($p_T > \Lambda/2$ GeV) & 1,250 & $ 8.0\times10^{-2}$ & $1.4\times10^{-1}$ & $6.4\times10^{-2}$ \\ 
 & 2,000 & $1.3\times10^{-2}$ & $9.5\times10^{-3}$ & $9.2\times10^{-3}$\\ \hline
\end{tabular}
\caption{Table of cross sections computed for this study for three scalar benchmarks. The inclusive $\phi$ production is taken from the experimental result~\cite{ATLAS,CMS:2015dxe}, while the background is the cross section from continuum diphoton production requiring $680 < m_{\gamma\gamma} < 820$ GeV and $p_T(\gamma\gamma) > \Lambda/2$ GeV. The bottom rows are the cross section for the given $\Lambda$ assuming the cross section from the first row and requiring that the $p_T$ of $\phi$ be greater than $\Lambda/2$ GeV. }
\label{tab:xsec}
\end{table}

In our analysis we set all couplings $c_i$ equal to one for simplicity. We note however, that because we are using normalized spectra, rescaling all $c$'s simultaneously will not affect the shape of the spectra. In other words, because the total production cross section of the $\phi$ is assumed to be known, this gives a measurement of $c$ so we do not need to care about it our analysis. Changing the size of $c_1$ relative to $c_2$ will change the results of the analysis, but we do not consider that here. This is discussed further in \sref{conc}. 

Having computed the cross section of signal plus background for a given $\Lambda$ with a $p_T$ cut of $\Lambda/2$, we can compare this to the cross section for very large $\Lambda$ plus background. We then want to estimate how much data it would take to discriminate between these two possibilities, which could in turn put a lower bound on the value of $\Lambda$. In order to compute this, we follow the hypothesis testing procedure described in, for example,~\cite{DeRujula:2010ys,Stolarski:2012ps}. The number of events for the two distributions is given by Poisson distributions around the expected number of events:
\begin{eqnarray}
&{\rm Poisson}\left[\mathcal{L}\times(\sigma(\Lambda) + \sigma({\rm BG} ))\right] \;\;\; &{\rm (Hypothesis \;1)} \nonumber\\
&{\rm Poisson}\left[\mathcal{L}\times(\sigma(\Lambda = 10\, {\rm TeV}) + \sigma({\rm BG}))\right] \;\;\; &{\rm (Hypothesis \;2)} 
\label{eq:poisson}
\end{eqnarray}
We can then use a general procedure for finding the overlap between two distributions, $f$ and $g$ which is to solve the equation
\begin{equation}
\int_0^{x_0} f(x) dx = \int_{x_0}^\infty g(x) dx
\label{eq:integral}
\end{equation}
for $x_0$. We have taken the mean of $f$ to be less than the mean of $g$.\footnote{The Poisson distribution is a discrete one over integer values, so the integrals in \eref{integral} should technically be sums, but as long as we are dealing with numbers of events $\gtrsim 10$, the continuous approximation is a good one.} Then the confidence level with which you can separate the two distributions is the value of that integral. This is shown in \fref{poisson} for the example of 50 and 74 events which can be separated at 95\% confidence. 

\begin{figure}[tb]
\centering
\includegraphics[width=0.5\textwidth]{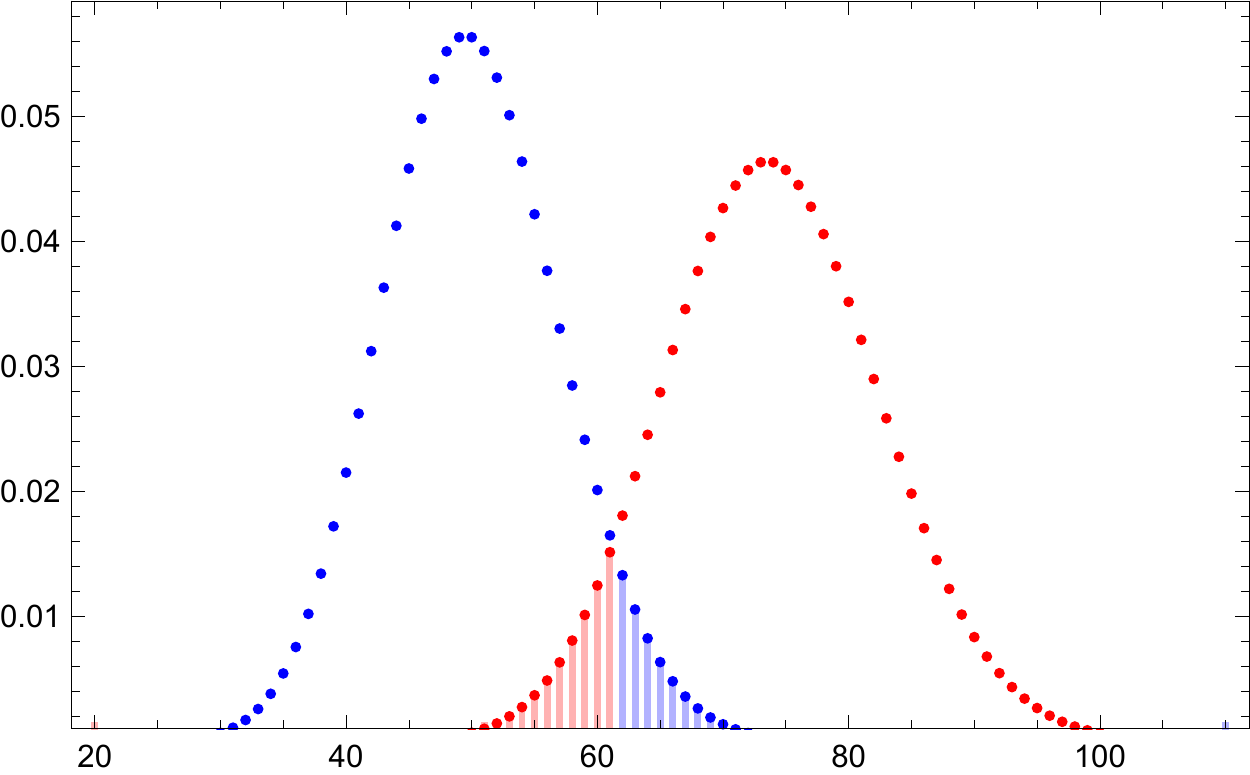}~~
\caption{Example of our statistical procedure of separating two Poisson distributions. Here the two distributions have mean of 50 (blue) and 74 (red), and the shaded area under both distriutions corresponds to 5\% probability, so these two hypotheses can be separated at 95\% confidence.  }
\label{fig:poisson}
\end{figure}

Given this procedure we can now determine how much luminosity is needed to separate the two hypotheses in \eref{poisson} as a function of $\Lambda$. This is shown in \fref{limit} for both the scalar and pseudo-scalar case. We also show the $\pm1\sigma$ uncertainty on this limit which is calculated by a one sigma upward (downward) fluctuation on hypothesis 1 and downward (upward) fluctuation on hypothesis 2, assuming that our data is distributed according to a Poisson distribution. The limit is shown as a function of luminosity, but it is normalized to the total number of $\phi$ events produced at the LHC. For the case of the 750 GeV resonance from~\cite{ATLAS, CMS:2015dxe}, the cross section used for normalization is 8 fb, but this analysis would apply to any resonance discovered in the future, although an adjustment may be nessesary if the signal to background ratio is significantly different.

\begin{figure}[tb]
\centering
\includegraphics[width=0.5\textwidth]{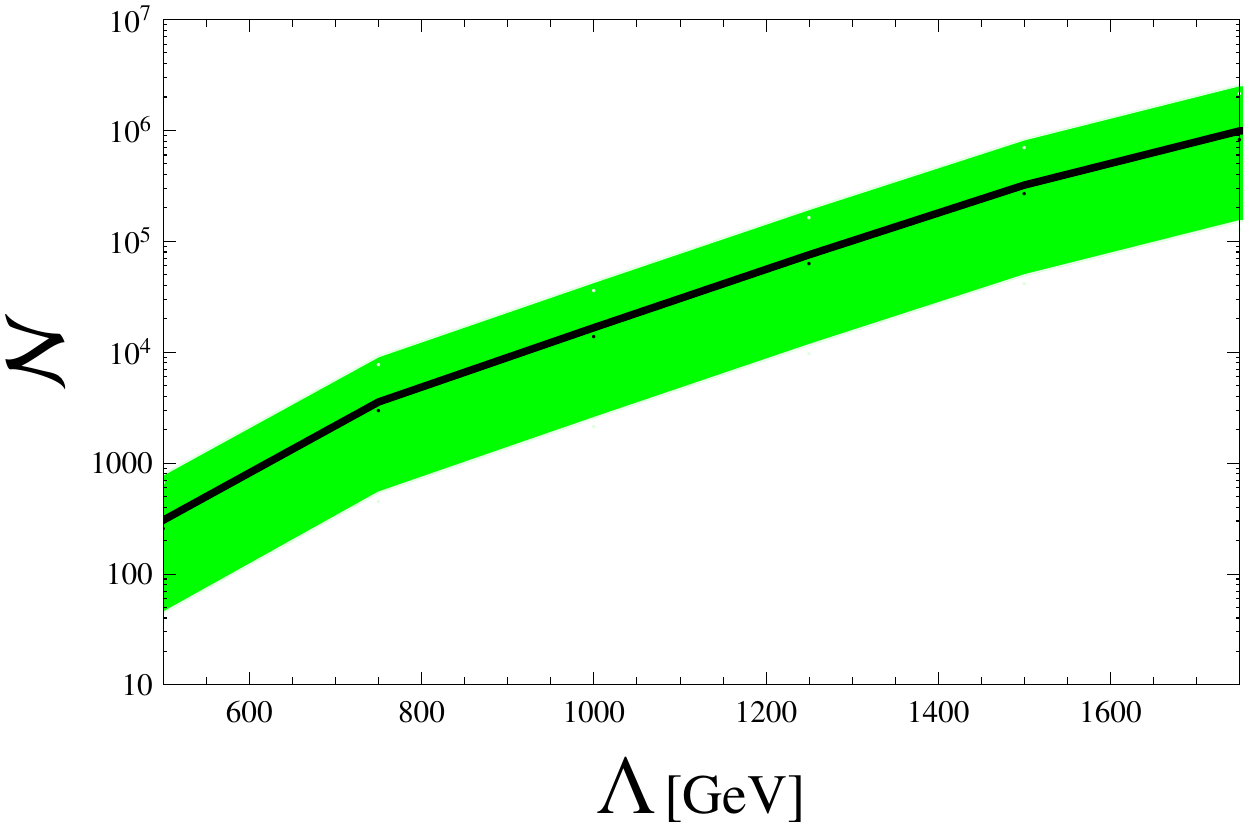}~~
\includegraphics[width=0.5\textwidth]{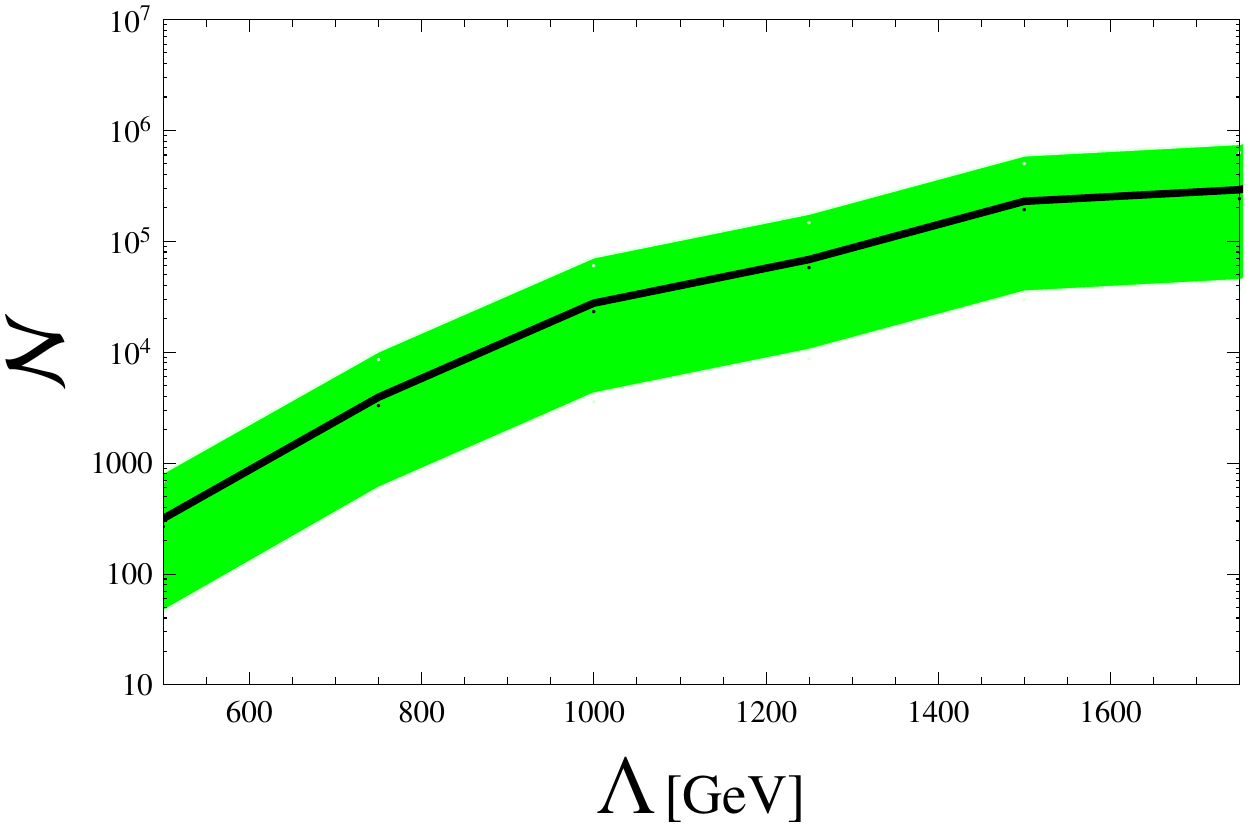}~~
\caption{Total number of $\phi$ events needed to rule out a given value of $\Lambda$ at 95\% denoted by the solid black line and the $\pm 1 \sigma$ expectation (green band) for a scalar (left) and psuedo-scalar (right). }
\label{fig:limit}
\end{figure}

We have performed this analysis for the scalar and pseudo-scalar cases, although the results are similar. This is because, to a reasonable approximation, the scaling of the cross section with $\Lambda$ can be computed by dimensional analysis, so this analysis is rather insensitive to the Lorentz structure of the vertex. The results for both cases are shown in \fref{limit}.

\fref{limit} shows us that with a relatively modest luminosity, enough to produce $\mathcal{O}(1000)$ $\phi$ events, one can put a lower bound on $\Lambda$ around the mass of the $\phi$ of 750 GeV. This will be particularly interesting if the resonance has a broad width as hinted by the ATLAS data~\cite{ATLAS}, which would point to strong coupling and a relatively low scale. In such a scenario, its possible this technique will fail to rule out specific values of $\Lambda$ when the actual limit differs significantly from the expected limit, in which case one could refine the statistical procedure to determine how to measure $\Lambda$ directly, but we leave this possibility to future work.

Our limit procedure involves treating the effective field theory as a good description all the way up to the kinematic limit of the collider, but at very high energy, especially for low $\Lambda$, the EFT is no longer an accurate description. There will, however, be some transition region where it is approximately correct. The regime of validity, or alternatively the accuracy of the EFT approximation, can be improved by including the dimension 7 operators as we do. Furthermore, in the case of a strongly interacting UV completion, naive dimensional analysis would indicate that the EFT is good up to a scale $\sim 4\pi\Lambda$. Finally, because of the rapidly falling parton distribution functions, most of the rate is coming from the region of phase space just above the $p_T$ cut. Therefore, even though some of our calculation is strictly outside of the regime of validity of the EFT, the bulk of the information is coming from the transition region, exactly the one we are trying to probe, and the EFT is the only way to do so model independently.

\section{Example UV Completion: Vector-like Quarks} \label{sec:vlq}

We are interested in probing the high energy behaviour of our putative resonance, and at high energy the effective field theory no longer becomes the correct description. We therefore want to understand how this analysis connects with a genuine UV completion for the effective field theory. As an example, we will take the well studied vector-like quark (VLQ) model or VolksModell\footnote{Everybody's Model.}~\cite{StumiaTalk}. The field content in addition to $\phi$ is a set of fermions $Q$ and $\bar{Q}$ which are colour triplets and vector-like under the SM gauge group. Then we add the Yukawa coupling
\begin{equation}
\lambda \, \phi \, \bar{Q} \, Q \, ,
\end{equation}
which, along with the gauge interactions of $Q$, is sufficient to generate the scalar higher dimension operators of \eref{op5} and \eref{op7}. There may be other dynamics that cause the $Q$ to decay, but that does not affect the phenomenology considered here. 

We take the $Q$ mass to be larger than $m_\phi/2$ so that on-shell decays of $\phi$ to $\bar{Q}Q$ are forbidden, allowing for a substantial branching fraction to photon pairs.  This model has the same structure as the SM Higgs coupling to the top quark, and that coupling of course is the dominant mediator of Higgs production via gluon fusion at the LHC. Therefore, we can compute the production of $\phi$ + jet using the SM formulae~\cite{Ellis:1987xu,Baur:1989cm} by making the translation
\begin{equation}
\frac{\alpha_W}{m_W^2} \rightarrow \frac{\lambda^2}{\pi m_Q^2},
\end{equation}
where $\alpha_W$ is the fine structure coupling for the $SU(2)$ gauge group of the SM. We compute the $p_T$ spectrum of the $\phi$ at parton level using the analytic results and requiring $|\eta_j| < 5$, $|\eta_\phi|< 2.5$ and $p_{T,\phi} > 100$ GeV. We do not simulate the decay of the $\phi$, so our simulation is not identical to the EFT case of \sref{eft}, but we will see that the approximations are reasonable. 

Because we are using normalized distributions, the value of $\lambda$ does not affect our analysis. This is analogous to how the overall scaling of the $c_i$'s in the EFT analysis is also unimportant. Therefore, we can compare the EFT and VLQ computations with the simple translation $\Lambda = m_Q$. When doing this comparison, we first note that the EFT and the UV completion should agree for large $\Lambda$ and heavy $m_Q$. As a test of our computation, we compare the $p_T$ distributions obtained from the EFT and from this UV completion in \fref{eft-compare}. We see that for large $\Lambda$ and $m_Q$ in the brown curves, that the two calculations agree completely. For intermediate $\Lambda = m_Q = 1250$ GeV, we see from the green curve that they agree at low $p_T$, but that the EFT over-estimates the cross section at high $p_T$. 
At low $m_Q$, the full loop effects of the VLQ's should be included to get a correct calculation and one is outside the regime of validity of the EFT. This is also evident from the red curves in \fref{eft-compare} corresponding to $\Lambda = m_Q = 750$ GeV, which look quite different for $p_T \gtrsim 600$ GeV.

\begin{figure}[tb]
\centering
\includegraphics[width=0.5\textwidth]{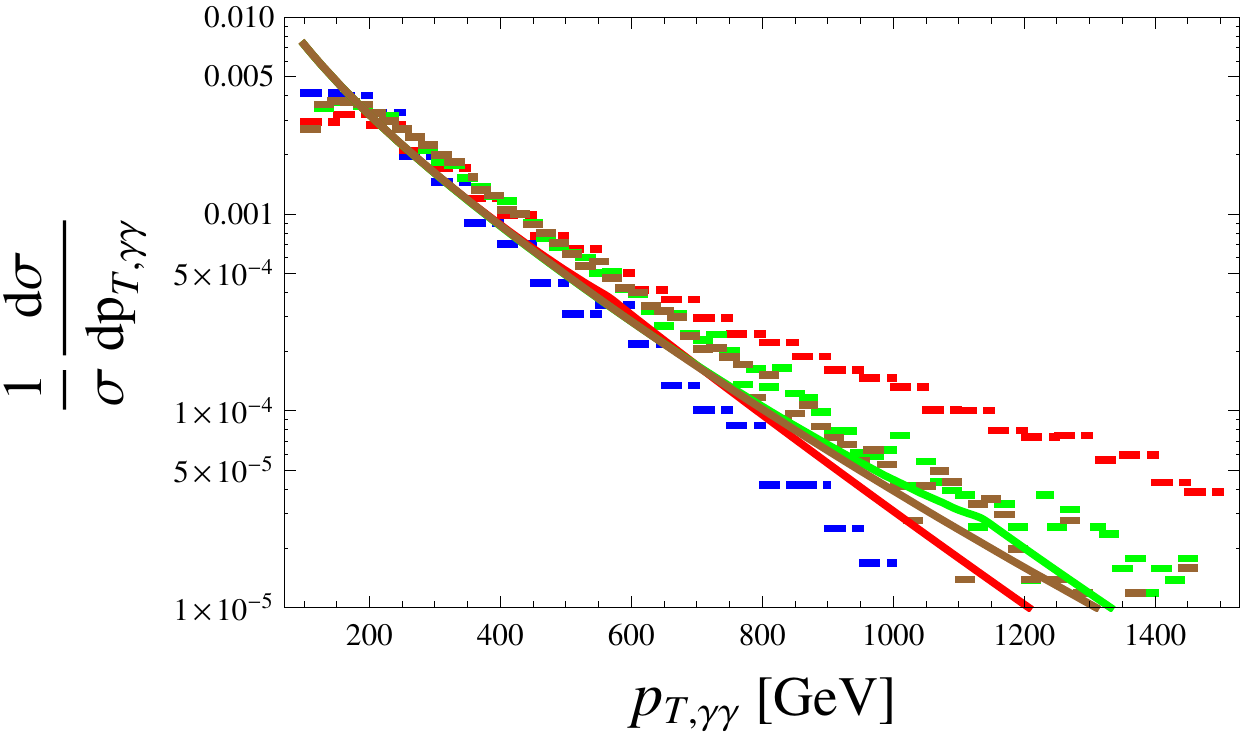}~~
\caption{Comparison of the $p_T$ spectra of the $\phi$ computed with the effective field theory, dashed histograms, and with the VLQ model (solid lines). The colours are as in \fref{eft-spectra}, $\Lambda = m_Q = 750$, 1250, and 2000 GeV in red, green, and brown, respectively. }
\label{fig:eft-compare}
\end{figure}

We now analyze how well the mass of underlying fermion in this particular UV completion is encoded in the $p_T$ spectrum. As we see in \fref{eft-compare}, the spectrum is still a rapidly falling function of $p_T$, even for very low values of $m_Q$. Therefore, we find it useful to look at ratios of normalized $p_T$ spectra, which are shown in \fref{uv-ratio}. Namely, we study the ratio of the $p_T$ spectrum for fixed $m_Q$ relative to that of $m_Q = 10$ TeV to compare a fixed mass to the EFT limit, analogous to the procedure used in \sref{eft}. For $p_T \gg m_Q$, we see that the full theory predicts a faster falling spectrum than the EFT limit as expected when the propagator of the internal fermion is dominated by momentum rather than mass at low energy. 

\begin{figure}[tb]
\centering
\includegraphics[width=0.5\textwidth]{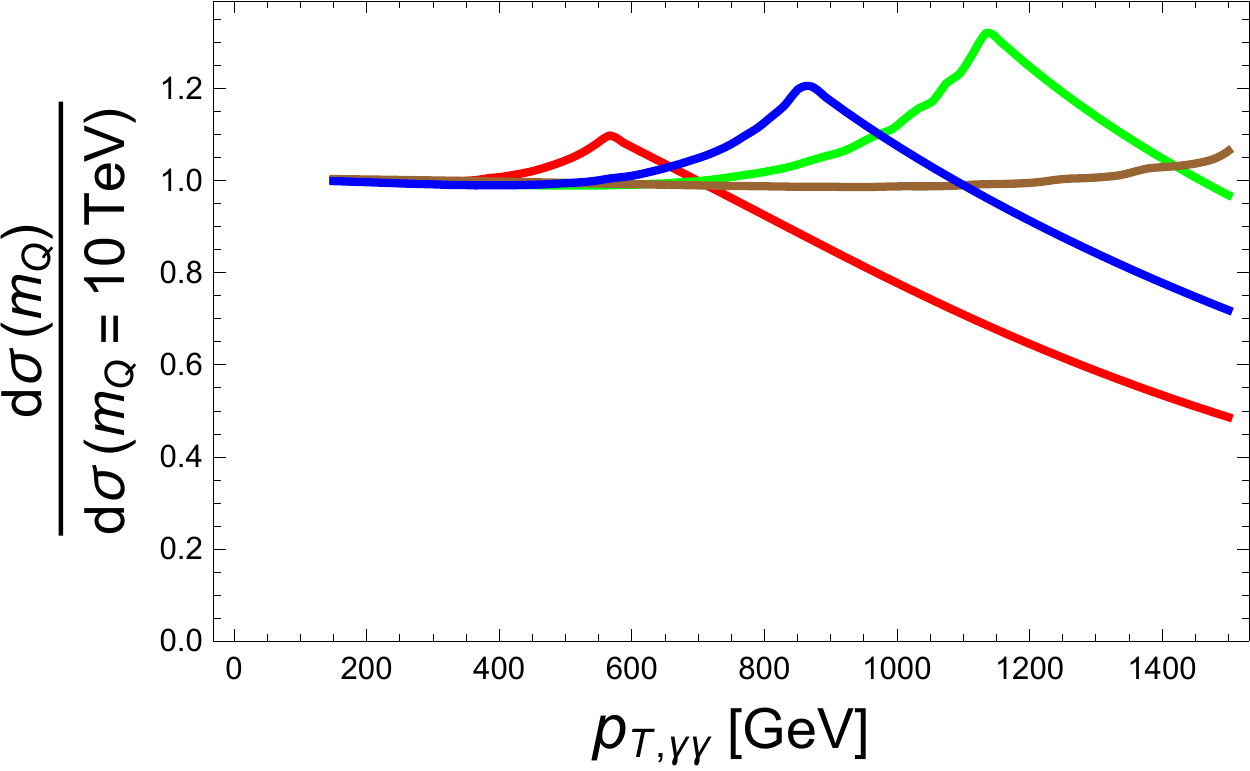}~~
\caption{Ratio of normalized $p_T$ distributions. The denominator is always the EFT limit of $m_Q = 10$ TeV.  The numerator is, going from bottom to top on the right side of the plot, $m_Q =$ 750, 1000, 1250, and 2000 GeV in red, blue, green, and brown, respectively. }
\label{fig:uv-ratio}
\end{figure}

\fref{uv-ratio} also shows that before the cross section begins to fall relative to the EFT, there is a rise and a peak at slightly below $m_Q$. This arises because of the $q\bar{q}$ initiated sub-process where all the energy of the process flows through a triangle loop of fermions. At $s=4m_Q^2$, the triangle loop function goes from being real to being complex and there is a discontinuity in the derivative. This does not occur in the $qg$ or $gg$ initiated processes because of the structure of the diagrams.\footnote{For the full set of diagrams contributing to this process see Fig.~1 of~\cite{Baur:1989cm}.} This behaviour can be seen in \fref{parton} where we plot the parton level cross section for the three different initial states as a function of $\sqrt{s}$. 

\begin{figure}[tb]
\centering
\includegraphics[width=0.5\textwidth]{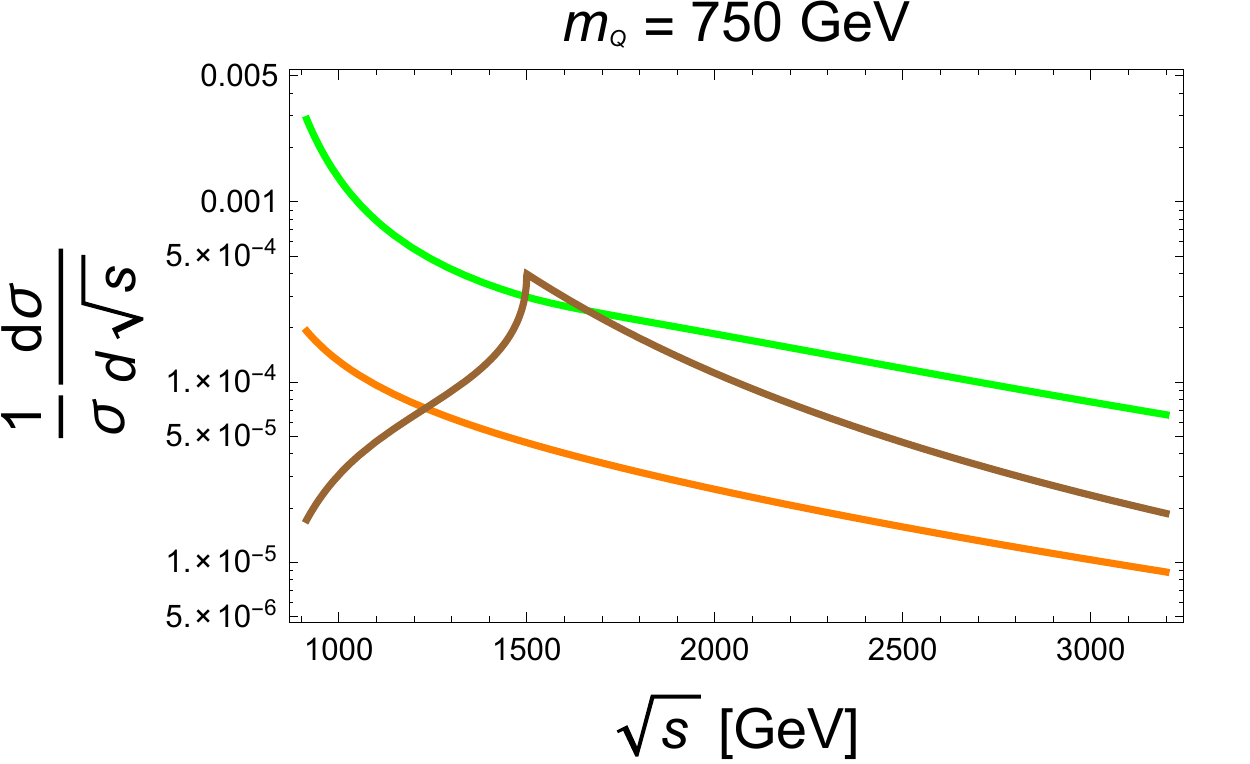}~~
\includegraphics[width=0.5\textwidth]{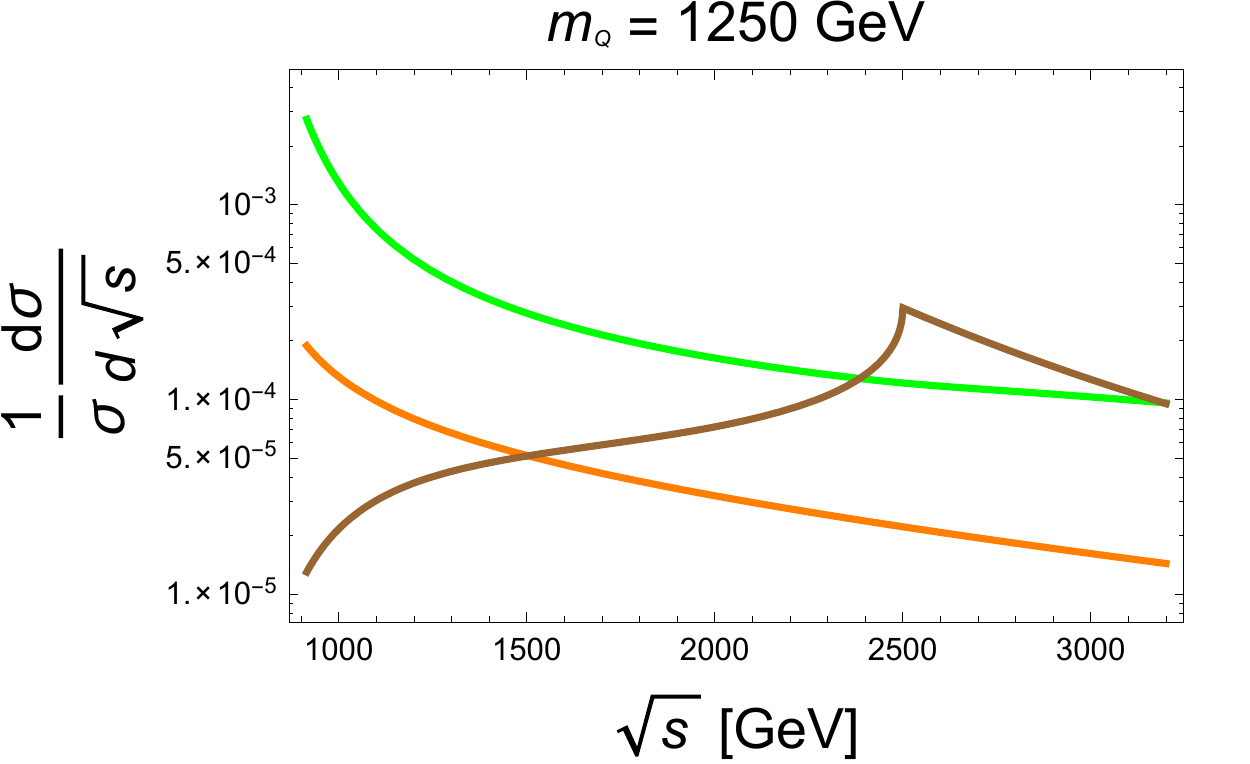}~~
\caption{Parton level cross section (no parton distribution functions) split up into the $gg$, $qg$ and $q\bar{q}$ initiated process in green, orange, and brown, respectively. On the left (right) side is $m_Q = 750$ GeV ($m_Q = 1250$ GeV). Here we have fixed the scattering angle to be $\theta = 0$. The curves are normalized such that the three integrals add to unity.  The range in $\sqrt{s}$ corresponds to approximately the same range in $p_T$ as shown in other spectrum plots.}
\label{fig:parton}
\end{figure}

We see that the total cross section is dominated by the gluon initial state and at low $s$. For relatively large masses, the very boosted regime uncovers the importance of the $q\bar{q}$ initiated process. If one were to get sufficient events, one could make out the peak and directly measure $m_Q$ should a model with this behaviour be the correct UV description. One possibility to enhance the signal would be to look at kinematic variables such as rapidity distributions and jet binning in order to discriminate $q\bar{q}$ initial states relative to the others~\cite{Gao:2015igz,Ebert:2016idf,Harland-Lang:2016vzm}. This could allow a more rapid discovery of the peak feature seen at $\sqrt{s} = 2 m_Q$ implying a precise measurement of the mass of the particle in the loop. 

One could apply the same statistical procedure from \sref{eft} to the $p_T$ spectrum for the UV completion, though care must be taken because there is a rise before the spectrum falls, so integration, namely counting all events above a $p_T$ cut, could wash out some of the differences between the EFT and the full model. The results will be similar to \fref{limit}, but further study is necessary to optimize the cuts in order to fully take advantage of shape of the distribution. On the other hand, the shape is specific to this type of model, and not necessarily generic, so we leave \fref{limit} in the effective field theory as our main result.

\section{Summary and Outlook} \label{sec:conc}

In this work, we have introduced a method to probe the underlying mechanism for a resonance produced from gluon initial states. In particular, we study events where a high $p_{T}$ jet is produced in association with the resonance. Our analysis focuses on a scalar or pseudoscalar resonance with a $750$ GeV mass that decays to photon pairs. For the case where effective field theory is a good description up to the scale $\Lambda$, we implemented a statistical method to derive lower bounds as a function of luminosity on the value of $\Lambda$. We see that for values of $\Lambda$ close to the TeV scale, one can probe these scenarios with as low as ${\cal O}\left(1000\right)$ events. We can also consider a UV completion for the model with vector-like quarks running in the production loop. This matches the EFT analysis quite well for heavy vector-like quarks, but new features emerge if the additional new states are at the TeV scale. 

Our analysis only accounts for statistical uncertainties in both the signal and the background, but a realistic analysis must also treat systematic uncertainties. One source of such uncertainties arises from the diphoton mass resolution. Typical values for the background are of $2$-$4\%$ at masses near $750$ GeV from a search by the ATLAS collaboration~\cite{Aaboud:2016tru} and roughly $5\%$ from a CMS search~\cite{Khachatryan:2015qba}. In addition, since we are looking at associated production of the resonance with a high $p_{T}$ jet, the determination of the absolute jet energy scale and resolution becomes also source of systematics. However, a recent study by the ATLAS collaboration on jet calibration and systematics at $\sqrt{s}=13$ TeV estimates this at the level of $1\%$ for jets with $p_{T}>200$ GeV~\cite{ATLASjets}. 

The most important systematic is from the uncertainty on the signal and background cross section which arises from using only tree level computations for the signal. For the background, we have used next-to-leading order cross section, and the signal to background ratio is relatively large, so this systematic will be subdominant. For the signal, however, this correction can be quite large, for example the NLO correction to inclusive $\phi$ production is $\mathcal{O}(50\%)$ for the gluon initial state~\cite{Franceschini:2015kwy,Bauer:2016lbe}. The calculations for Higgs + jet exist at next-to-next-to leading order~\cite{Boughezal:2013uia}, so there should be no obstruction to extending those to a new resonance should it be discovered. The method developed in this work can be straightforwardly modified to include systematics, but most of the systematics are small or can be kept under control. 

In addition, the possibility of measuring the value of $\Lambda$ rather than placing a lower limit based on the amount of data collected at a high energy collider is worth studying further. One may in principle carry out a bin by bin analysis of the target distribution and maximize a likelihood function based on Poisson distributed number of events in each bin to extract the value of $\Lambda$ that best fits the data. This is especially useful if the scale $\Lambda$ is low and then ultimately a very precise measurement will be possible, but we leave the details to future work. 

In this analysis we have assumed a spin-0 state but this does not qualitatively affect our results. We can see from comparing scalar and pseudo-scalar in \fref{limit} that the Lorentz quantum numbers are not very important. The scaling with cross section for different values of $\Lambda$ is the dominant effect, and that can be derived essentially from dimensional analysis. 

One can also imagine changing the size of the EFT couplings, $c_i$. If they are all reduced by the same amount, then the cross section simply decreases, but all normalized distributions used in this study remain unchanged. If one increases $c_2$ relative to $c_1$, then the sensitivity to the scale $\Lambda$ will improve. The relative sizes of the $c's$ are a priori unknown, but one could imagine studying this further in particular UV theories where it can be predicted.

The search program for physics beyond the Standard Model is on. Even though as yet no new physics has been discovered, we have taken the optimistic approach of preparing for a new discovery, focusing on a resonance that couples to gluons. Such a resonance is special because it must be described by a non-renormalizble theory or a loop process. Here we have shown that the structure of the interaction with gluons can be uncovered with sufficient data, independent of the nature of the states that mediate these couplings. We hope these techniques will be applied to a newly discovered resonance in the near future.

\section*{Acknowledgements}
We would like to thank Zackaria Chacko, David Morrissey, Ann Nelson, and Roberto Vega-Morales for valuable feedback. This work was supported in part by the Natural Sciences and Engineering Research Council of Canada (NSERC).



\begin{thebibliography}{99}


\bibitem{ATLAS} 
  The ATLAS collaboration,
  ``Search for resonances decaying to photon pairs in 3.2 fb$^{-1}$ of $pp$ collisions at $\sqrt{s}$ = 13 TeV with the ATLAS detector,''
  ATLAS-CONF-2015-081.

\bibitem{CMS:2015dxe} 
  CMS Collaboration [CMS Collaboration],
  ``Search for new physics in high mass diphoton events in proton-proton
  collisions at 13TeV,''
  CMS-PAS-EXO-15-004.
 
\bibitem{Strumia:2016wys} 
  A.~Strumia,
  arXiv:1605.09401 [hep-ph].
  
\bibitem{Knapen:2015dap} 
  S.~Knapen, T.~Melia, M.~Papucci and K.~Zurek,
  Phys.\ Rev.\ D {\bf 93}, no. 7, 075020 (2016)
  doi:10.1103/PhysRevD.93.075020
  [arXiv:1512.04928 [hep-ph]].
  
\bibitem{Berthier:2015vbb} 
  L.~Berthier, J.~M.~Cline, W.~Shepherd and M.~Trott,
  JHEP {\bf 1604}, 084 (2016)
  doi:10.1007/JHEP04(2016)084
  [arXiv:1512.06799 [hep-ph]].
  
\bibitem{Kamenik:2016tuv} 
  J.~F.~Kamenik, B.~R.~Safdi, Y.~Soreq and J.~Zupan,
  arXiv:1603.06566 [hep-ph].
  
\bibitem{Franceschini:2016gxv} 
  R.~Franceschini, G.~F.~Giudice, J.~F.~Kamenik, M.~McCullough, F.~Riva, A.~Strumia and R.~Torre,
  arXiv:1604.06446 [hep-ph].

\bibitem{Harigaya:2015ezk} 
  K.~Harigaya and Y.~Nomura,
  Phys.\ Lett.\ B {\bf 754}, 151 (2016)
  doi:10.1016/j.physletb.2016.01.026
  [arXiv:1512.04850 [hep-ph]].

\bibitem{Nakai:2015ptz} 
  Y.~Nakai, R.~Sato and K.~Tobioka,
  Phys.\ Rev.\ Lett.\  {\bf 116}, no. 15, 151802 (2016)
  doi:10.1103/PhysRevLett.116.151802
  [arXiv:1512.04924 [hep-ph]].
  
\bibitem{Franceschini:2015kwy} 
  R.~Franceschini {\it et al.},
  JHEP {\bf 1603}, 144 (2016)
  doi:10.1007/JHEP03(2016)144
  [arXiv:1512.04933 [hep-ph]].

\bibitem{Molinaro:2015cwg} 
  E.~Molinaro, F.~Sannino and N.~Vignaroli,
  arXiv:1512.05334 [hep-ph].

\bibitem{Bai:2015nbs} 
  Y.~Bai, J.~Berger and R.~Lu,
  Phys.\ Rev.\ D {\bf 93}, no. 7, 076009 (2016)
  doi:10.1103/PhysRevD.93.076009
  [arXiv:1512.05779 [hep-ph]].

\bibitem{Belyaev:2015hgo} 
  A.~Belyaev, G.~Cacciapaglia, H.~Cai, T.~Flacke, A.~Parolini and H.~SerÙdio,
  arXiv:1512.07242 [hep-ph].

\bibitem{Bian:2015kjt} 
  L.~Bian, N.~Chen, D.~Liu and J.~Shu,
  Phys.\ Rev.\ D {\bf 93}, no. 9, 095011 (2016)
  doi:10.1103/PhysRevD.93.095011
  [arXiv:1512.05759 [hep-ph]].

\bibitem{Craig:2015lra} 
  N.~Craig, P.~Draper, C.~Kilic and S.~Thomas,
  Phys.\ Rev.\ D {\bf 93}, no. 11, 115023 (2016)
  doi:10.1103/PhysRevD.93.115023
  [arXiv:1512.07733 [hep-ph]].

\bibitem{Franzosi:2016wtl} 
  D.~Buarque Franzosi and M.~T.~Frandsen,
  arXiv:1601.05357 [hep-ph].

\bibitem{Harigaya:2016pnu} 
  K.~Harigaya and Y.~Nomura,
  JHEP {\bf 1603}, 091 (2016)
  doi:10.1007/JHEP03(2016)091
  [arXiv:1602.01092 [hep-ph]].
 
\bibitem{Hong:2016uou} 
  D.~K.~Hong and D.~H.~Kim,
  Phys.\ Lett.\ B {\bf 758}, 370 (2016)
  doi:10.1016/j.physletb.2016.05.034
  [arXiv:1602.06628 [hep-ph]].

\bibitem{Redi:2016kip} 
  M.~Redi, A.~Strumia, A.~Tesi and E.~Vigiani,
  JHEP {\bf 1605}, 078 (2016)
  doi:10.1007/JHEP05(2016)078
  [arXiv:1602.07297 [hep-ph]].

\bibitem{Harigaya:2016eol} 
  K.~Harigaya and Y.~Nomura,
  arXiv:1603.05774 [hep-ph].
  
\bibitem{Kamenik:2016izk} 
  J.~F.~Kamenik and M.~Redi,
  doi:10.1016/j.physletb.2016.06.060
  arXiv:1603.07719 [hep-ph].

\bibitem{Ko:2016sht} 
  P.~Ko, C.~Yu and T.~C.~Yuan,
  arXiv:1603.08802 [hep-ph].

\bibitem{Foot:2016llc} 
  R.~Foot and J.~Gargalionis,
  arXiv:1604.06180 [hep-ph].

\bibitem{Iwamoto:2016ral} 
  S.~Iwamoto, G.~Lee, Y.~Shadmi and R.~Ziegler,
  arXiv:1604.07776 [hep-ph].
  
\bibitem{Hamaguchi:2016umx} 
  K.~Hamaguchi and S.~P.~Liew,
  arXiv:1604.07828 [hep-ph].
  
\bibitem{Matsuzaki:2016joz} 
  S.~Matsuzaki and K.~Yamawaki,
  Phys.\ Rev.\ D {\bf 93}, no. 11, 115027 (2016)
  doi:10.1103/PhysRevD.93.115027
  [arXiv:1605.04667 [hep-ph]].
  
\bibitem{Bai:2016vca} 
  Y.~Bai, J.~Berger, J.~Osborne and B.~A.~Stefanek,
  arXiv:1605.07183 [hep-ph].
  
  
\bibitem{Harlander:2013oja} 
  R.~V.~Harlander and T.~Neumann,
  ``Probing the nature of the Higgs-gluon coupling,''
  Phys.\ Rev.\ D {\bf 88}, 074015 (2013)
  doi:10.1103/PhysRevD.88.074015
  [arXiv:1308.2225 [hep-ph]].
  
\bibitem{Grojean:2013nya} 
  C.~Grojean, E.~Salvioni, M.~Schlaffer and A.~Weiler,
  JHEP {\bf 1405}, 022 (2014)
  doi:10.1007/JHEP05(2014)022
  [arXiv:1312.3317 [hep-ph]].
  
\bibitem{Banfi:2013yoa} 
  A.~Banfi, A.~Martin and V.~Sanz,
  JHEP {\bf 1408}, 053 (2014)
  doi:10.1007/JHEP08(2014)053
  [arXiv:1308.4771 [hep-ph]].

\bibitem{Azatov:2013xha} 
  A.~Azatov and A.~Paul,
  JHEP {\bf 1401}, 014 (2014)
  doi:10.1007/JHEP01(2014)014
  [arXiv:1309.5273 [hep-ph]].

\bibitem{Bramante:2014gda} 
  J.~Bramante, A.~Delgado and A.~Martin,
  Phys.\ Rev.\ D {\bf 89}, no. 9, 093006 (2014)
  doi:10.1103/PhysRevD.89.093006
  [arXiv:1402.5985 [hep-ph]].

\bibitem{Schlaffer:2014osa} 
  M.~Schlaffer, M.~Spannowsky, M.~Takeuchi, A.~Weiler and C.~Wymant,
  Eur.\ Phys.\ J.\ C {\bf 74}, no. 10, 3120 (2014)
  doi:10.1140/epjc/s10052-014-3120-z
  [arXiv:1405.4295 [hep-ph]].

\bibitem{Buschmann:2014twa} 
  M.~Buschmann, C.~Englert, D.~Goncalves, T.~Plehn and M.~Spannowsky,
  Phys.\ Rev.\ D {\bf 90}, no. 1, 013010 (2014)
  doi:10.1103/PhysRevD.90.013010
  [arXiv:1405.7651 [hep-ph]].

\bibitem{Dawson:2014ora} 
  S.~Dawson, I.~M.~Lewis and M.~Zeng,
  Phys.\ Rev.\ D {\bf 90}, no. 9, 093007 (2014)
  doi:10.1103/PhysRevD.90.093007
  [arXiv:1409.6299 [hep-ph]].

\bibitem{Ghosh:2014wxa} 
  D.~Ghosh and M.~Wiebusch,
  Phys.\ Rev.\ D {\bf 91}, no. 3, 031701 (2015)
  doi:10.1103/PhysRevD.91.031701
  [arXiv:1411.2029 [hep-ph]].

\bibitem{Edezhath:2015lga} 
  R.~Edezhath,
  arXiv:1501.00992 [hep-ph].

\bibitem{Dawson:2015gka} 
  S.~Dawson, I.~M.~Lewis and M.~Zeng,
  Phys.\ Rev.\ D {\bf 91}, 074012 (2015)
  doi:10.1103/PhysRevD.91.074012
  [arXiv:1501.04103 [hep-ph]].

\bibitem{Langenegger:2015lra} 
  U.~Langenegger, M.~Spira and I.~Strebel,
  arXiv:1507.01373 [hep-ph].
  
\bibitem{Bishara:2016jga} 
  F.~Bishara, U.~Haisch, P.~F.~Monni and E.~Re,
  arXiv:1606.09253 [hep-ph].
  
\bibitem{Soreq:2016rae} 
  Y.~Soreq, H.~X.~Zhu and J.~Zupan,
  arXiv:1606.09621 [hep-ph].
  
\bibitem{Gracey:2002he} 
  J.~A.~Gracey,
  Nucl.\ Phys.\ B {\bf 634}, 192 (2002)
  Erratum: [Nucl.\ Phys.\ B {\bf 696}, 295 (2004)]
  doi:10.1016/S0550-3213(02)00334-6, 10.1016/j.nuclphysb.2004.06.053
  [hep-ph/0204266].
  
\bibitem{Neill:2009tn} 
  D.~Neill,
  arXiv:0908.1573 [hep-ph].
  


\bibitem{Alloul:2013bka} 
  A.~Alloul, N.~D.~Christensen, C.~Degrande, C.~Duhr and B.~Fuks,
  ``FeynRules  2.0 - A complete toolbox for tree-level phenomenology,''
  Comput.\ Phys.\ Commun.\  {\bf 185}, 2250 (2014)
  doi:10.1016/j.cpc.2014.04.012
  [arXiv:1310.1921 [hep-ph]].
  
\bibitem{Alwall:2011uj} 
  J.~Alwall, M.~Herquet, F.~Maltoni, O.~Mattelaer and T.~Stelzer,
  ``MadGraph 5 : Going Beyond,''
  JHEP {\bf 1106}, 128 (2011)
  doi:10.1007/JHEP06(2011)128
  [arXiv:1106.0522 [hep-ph]].

\bibitem{Sjostrand:2006za} 
  T.~Sjostrand, S.~Mrenna and P.~Z.~Skands,
  ``PYTHIA 6.4 Physics and Manual,''
  JHEP {\bf 0605}, 026 (2006)
  doi:10.1088/1126-6708/2006/05/026
  [hep-ph/0603175].
  
\bibitem{deFavereau:2013fsa} 
  J.~de Favereau {\it et al.} [DELPHES 3 Collaboration],
  ``DELPHES 3, A modular framework for fast simulation of a generic collider experiment,''
  JHEP {\bf 1402}, 057 (2014)
  doi:10.1007/JHEP02(2014)057
  [arXiv:1307.6346 [hep-ex]].


\bibitem{Gehrmann:2013aga} 
  T.~Gehrmann, N.~Greiner and G.~Heinrich,
  ``Photon isolation effects at NLO in $\gamma \gamma$ + jet final states in hadronic collisions,''
  JHEP {\bf 1306}, 058 (2013)
  Erratum: [JHEP {\bf 1406}, 076 (2014)]
  doi:10.1007/JHEP06(2014)076, 10.1007/JHEP06(2013)058
  [arXiv:1303.0824 [hep-ph]].

\bibitem{DeRujula:2010ys} 
  A.~De Rujula, J.~Lykken, M.~Pierini, C.~Rogan and M.~Spiropulu,
  Phys.\ Rev.\ D {\bf 82}, 013003 (2010)
  doi:10.1103/PhysRevD.82.013003
  [arXiv:1001.5300 [hep-ph]].
  
\bibitem{Stolarski:2012ps} 
  D.~Stolarski and R.~Vega-Morales,
  Phys.\ Rev.\ D {\bf 86}, 117504 (2012)
  doi:10.1103/PhysRevD.86.117504
  [arXiv:1208.4840 [hep-ph]].
  
\bibitem{StumiaTalk}
  Talk by A.~Strumia at the Moriond 2016 conference. 
  \verb!https://indico.in2p3.fr/event/12279/session/12/contribution/110/material/slides/1.pdf!  
  
\bibitem{Ellis:1987xu} 
  R.~K.~Ellis, I.~Hinchliffe, M.~Soldate and J.~J.~van der Bij,
  Nucl.\ Phys.\ B {\bf 297}, 221 (1988).
  doi:10.1016/0550-3213(88)90019-3
  
\bibitem{Baur:1989cm} 
  U.~Baur and E.~W.~N.~Glover,
  Nucl.\ Phys.\ B {\bf 339}, 38 (1990).
  doi:10.1016/0550-3213(90)90532-I
  
\bibitem{Gao:2015igz} 
  J.~Gao, H.~Zhang and H.~X.~Zhu,
  Eur.\ Phys.\ J.\ C {\bf 76}, no. 6, 348 (2016)
  doi:10.1140/epjc/s10052-016-4200-z
  [arXiv:1512.08478 [hep-ph]].
  
\bibitem{Ebert:2016idf} 
  M.~A.~Ebert, S.~Liebler, I.~Moult, I.~W.~Stewart, F.~J.~Tackmann, K.~Tackmann and L.~Zeune,
  arXiv:1605.06114 [hep-ph].
  
\bibitem{Harland-Lang:2016vzm} 
  L.~A.~Harland-Lang, V.~A.~Khoze, M.~G.~Ryskin and M.~Spannowsky,
  arXiv:1606.04902 [hep-ph].
  
\bibitem{Aaboud:2016tru} 
  M.~Aaboud {\it et al.} [ATLAS Collaboration],
  arXiv:1606.03833 [hep-ex].
  
\bibitem{Khachatryan:2015qba} 
  V.~Khachatryan {\it et al.} [CMS Collaboration],
  Phys.\ Lett.\ B {\bf 750}, 494 (2015)
  doi:10.1016/j.physletb.2015.09.062
  [arXiv:1506.02301 [hep-ex]].
  
  \bibitem{ATLASjets}
  The ATLAS Collaboration,
  ATL-PHYS-PUB-2015-015
  
\bibitem{Bauer:2016lbe} 
  M.~Bauer, C.~Hoerner and M.~Neubert,
  arXiv:1603.05978 [hep-ph].
  
\bibitem{Boughezal:2013uia} 
  R.~Boughezal, F.~Caola, K.~Melnikov, F.~Petriello and M.~Schulze,
  JHEP {\bf 1306}, 072 (2013)
  doi:10.1007/JHEP06(2013)072
  [arXiv:1302.6216 [hep-ph]].
  
  
  
  

  
  
  
  
  
\end{thebibliography}

\end{document}